\documentclass[conference]{IEEEtran}

\makeatletter

\usepackage[T1]{fontenc}
\usepackage[utf8]{inputenc}
\usepackage[switch]{lineno} 

\usepackage{amsmath}
\usepackage{amssymb}
\usepackage{array}
\usepackage{algorithm}
\usepackage{algorithmic}

\usepackage{cite}
\usepackage{graphicx}
\usepackage{listings}
\usepackage{subcaption}
\usepackage{multirow}

\usepackage{enumitem}
\usepackage[dvipsnames, table]{xcolor}

\usepackage{booktabs} 
\usepackage{siunitx}  
\usepackage{multirow} 
\usepackage{url}
\begin{document}

\title{Cross-Domain Query Translation for Network Troubleshooting: A Multi-Agent LLM Framework with Privacy Preservation and Self-Reflection}

\author{\IEEEauthorblockN{Nguyen Phuc Tran and Brigitte Jaumard}
\IEEEauthorblockA{  \textit{Computer Science and Software Engineering} \\
                    \textit{Concordia University} \\
                    Montréal (Qc) Canada \\
                    brigitte.jaumard@concordia.ca}
\and
\IEEEauthorblockN{Karthikeyan Premkumar and Salman Memon}
\IEEEauthorblockA{  \textit{Ericsson Montréal} \\
                    Montréal (Qc) Canada
                 }
}

\maketitle

\begin{abstract}

This paper presents a hierarchical multi-agent LLM architecture to bridge communication gaps between non-technical end users and telecommunications domain experts in private network environments.
We propose a cross-domain query translation framework that leverages specialized language models coordinated through multi-agent reflection-based reasoning. 
The resulting system addresses three critical challenges: 
(1) accurately classify user queries related to telecommunications network issues using a dual-stage hierarchical approach, 
(2) preserve user privacy through the anonymization of semantically relevant personally identifiable information (PII) while maintaining diagnostic utility, and 
(3) translate technical expert responses into user-comprehensible language. 

Our approach employs ReAct-style agents enhanced with self-reflection mechanisms for iterative output refinement, semantic-preserving anonymization techniques respecting $k$-anonymity and differential privacy principles, and few-shot learning strategies designed for limited training data scenarios. 
The framework was comprehensively evaluated on 10,000 previously unseen validation scenarios across various vertical industries.

\end{abstract}



\section{Introduction}

Private networks deployed in vertical sectors such as hospital telemetry systems, factory automation, and autonomous vehicles have become critical infrastructure components.
These networks rely on contemporary communication technologies, including 4G and 5G, and are increasingly supported by customer-facing LLM-based assistants that enhance user interaction~\cite{Ahokangas_2029}.
However, when network problems occur, users often lack the specialized vocabulary needed to describe them in terms that are readily interpretable by network operators or telecommunications-focused LLM assistants.
Moreover, in-depth analysis requires vendor-specific knowledge and topology information that organizations cannot readily disclose because of privacy, security, and regulatory constraints.
At the same time, users require explanations in plain language so that they can participate effectively in troubleshooting.

For example, a user may report: "My screen has red dots blinking, and response is very slow when uploading data".
From a telecommunications perspective, this description may indicate radio resource contention, packet loss, elevated latency, or cell-reselection problems. Network professionals would instead examine key performance indicators (KPIs), signal-quality measurements, device configurations, and interference patterns.
Directly forwarding such observations to a specialist telecom LLM raises several challenges:

\begin{enumerate}

\item \textbf{Query Semantic Gap}: Users employ visual or behavioral terminology, whereas telecom experts rely on network measurements and protocol-specific vocabulary.

\item \textbf{Privacy Constraints}: User communications may contain sensitive information, such as personal identifiers, location data, network topology, and device addresses.

\item \textbf{Data Governance}: Organizational policies prevent exchanging raw user questions with outside telecom specialists.

\item \textbf{Response Comprehensibility}: Experts use technical terminology (e.g., "PRB utilization exceeds 80\%"), which must be translated for non-technical users.
\end{enumerate}

To address these challenges, this paper presents a multilayer hierarchical framework composed of cooperating specialized LLM agents. The main contributions are as follows:

\begin{itemize}
\item \textbf{Two-Stage Hierarchical Classification}: A combination of lightweight machine-learning models and LLM-based semantic reasoning for accurate telecom-query identification and intent extraction with limited labeled data~\cite{Tunstall_2022, brown2020language}.

\item \textbf{Semantic-Preserving Anonymization}: A privacy-conscious transformation technique that applies context-aware entity detection and semantic criticality assessment to mask personally identifiable information (PII) while preserving diagnostic utility and supporting differential privacy and $k$-anonymity principles~\cite{Sweeney_2002}.

\item \textbf{Reflection-Augmented Agent Coordination}: A multi-agent environment based on ReAct reasoning patterns~\cite{Yao_2023}, reinforced with self-reflection loops to refine outputs, detect hallucinations, and correct errors without human supervisory intervention~\cite{Shinn_2023, Madaan_2023}.

\item \textbf{Few-Shot Domain Adaptation}: Prompt optimization under low-resource conditions through example selection and in-context learning~\cite{Wei_2022, Reimers_2019}.

\item \textbf{Comprehensive Evaluation on 10,000 Scenarios}: Validation on a diverse set of previously unseen query-response scenarios spanning multiple vertical domains and failure categories, together with a detailed analysis of the resulting performance trends.
\end{itemize}

The remainder of the paper is organized as follows. Section~\ref{sec:related} reviews related work. Section~\ref{sec:architecture-methodology} describes the proposed architecture, its components, and the underlying implementation choices, including the classification algorithms, anonymization pipeline, and agent-coordination strategy. Section~\ref{sec:evaluation} presents the evaluation on 10,000 previously unseen validation scenarios. Section~\ref{sec:discussion} concludes the paper and outlines limitations and future work.

\section{Related Work}\label{sec:related}

Domain adaptation for specialized NLP tasks is challenged by data scarcity and distribution shift.
Prior work explores domain-adaptive pre-training and few-shot in-context learning~\cite{Ramponi_2020, devlin2019bert}, as well as structured prompting and CoT-based intent detection~\cite{zhou2025llm}.
These methods improve generalization with limited labels but remain sensitive to domain mismatch, motivating architectures that incorporate explicit reasoning and cross-domain guidance.

\subsection{Privacy Protection in AI Applications}

Privacy-preserving NLP commonly relies on differential privacy, federated learning, and PII masking~\cite{he2024emerged, bonawitz2021federated, habernal2023privacy, mahendran2021privacy}.
However, telecom diagnostics depend on relational patterns, topology, device linkage, and temporal behavior, which naive masking can destroy.
This gap highlights the need for anonymization mechanisms that balance structural fidelity with strong privacy guarantees.

These observations point to an opportunity for new methods in cross-domain translation for telecom networks.
The current literature offers limited support for explicit cross-agent validation or integration with external verification mechanisms, particularly when self-evaluation is unreliable. A framework that incorporates such verification can therefore address an important open problem in this domain.


\subsection{Coordination and Domain-Specific Adaptation in Telecom}

Telecommunications requires domain-adapted LLMs with strict verification due to specialized terminology, compliance constraints, and the need for diagnostically reliable outputs.
While domain-specific pretraining and telecom-oriented RAG systems show improvements~\cite{zou2025telecomgpt, Bornea_2024}, coordination and verification across agents remain underexplored. This motivates a unified framework that integrates multi-agent reasoning, privacy preservation, and cross-domain translation.


\section{System Architecture and Methodology}\label{sec:architecture-methodology}


\subsection{Architectural Overview}

\begin{figure}[htb]
    \centering
    \includegraphics[width=1\linewidth]{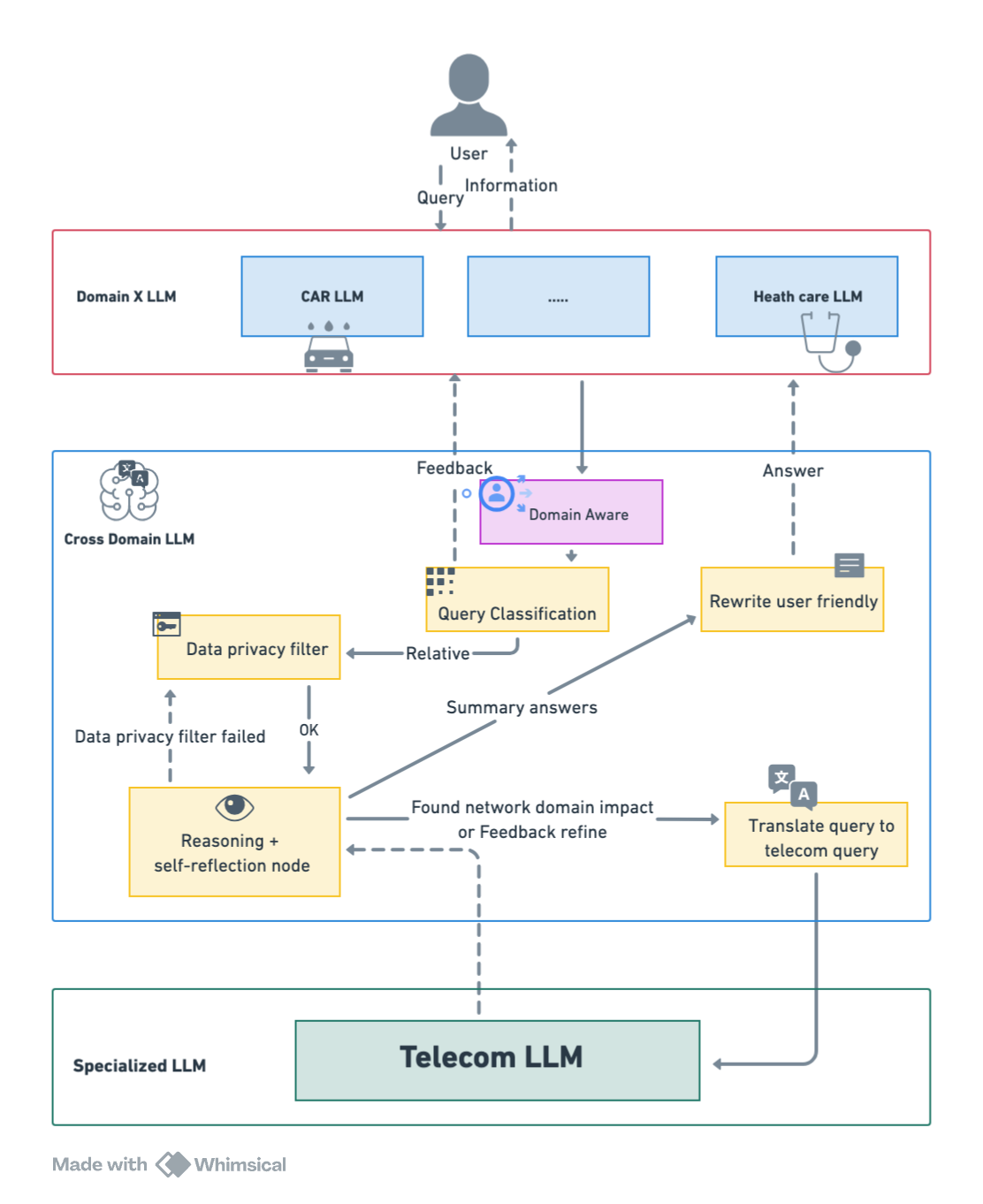}
    \caption{Proposed Cross-Domain Query Translation Framework.}
    \label{fig:proposal-cross-domain}
\end{figure}
The proposed cross-domain query translation framework uses a hierarchical multi-agent design.
Specialized agents work together on specific subtasks. 
They communicate through clear interfaces that define message formats and data schemas~\cite{li2024survey}. 
Fig.~\ref{fig:proposal-cross-domain} shows the system built from three connected layers:
\begin{enumerate}
    \item \textit{Customer Interaction Layer}: This layer takes queries from end-users or other LLMs and links to the cross-domain LLM layer with standard protocols across sectors.

    \item \textit{Cross-Domain LLM Layer}: This layer connects customer-facing LLMs to telecom LLMs. It includes domain-aware query classification, privacy filtering, translation of user queries into telecom terms, response simplification, and a reasoning module that self-reflects on sub-component outputs.

    \item \textit{Specialized LLM Layer}: This layer talks directly to telecom LLMs to create technical diagnoses and recommendations. It receives the translated query from the cross-domain layer plus any feedback from self-reflection.
\end{enumerate}
The layered design keeps components separate.
As a result, it simplifies the development and maintenance of each component without disrupting the overall system.
It also supports the addition of new features and flexible scaling across environments through the LangGraph framework\footnote{\url{https://www.langchain.com/langgraph}} with a router.
Furthermore, prompts and evaluation criteria within the cross-domain LLM layer are managed with LangFuse\footnote{\url{https://langfuse.com/}}, enabling dynamic control of prompts and criteria for each subcomponent.


\subsection{Component Specification and Detailed Methodology}


\subsubsection{Domain-Aware Component}

This component translates query distributions to industrial domains using a soft classification strategy that includes LLM-based categorization and confidence scoring.
For each query, it outputs a probability vector that shows domain likelihoods.
Then, it adds confidence scores based on entropy and margin values.
These outputs make domain decisions easier to understand and help prioritize contexts for later reasoning.
The system relies on structured taxonomies and feature embeddings that capture industries, device operations, and past patterns~\cite{minaee2021deep}.
Thus, it reduces misclassification and works with diverse scenarios.
In addition, the component also includes prompt engineering modules to build domain-specific instructions and prompts from embeddings and templates.
This step uses the reasoning and self-reflection method to refine itself.
Consequently, it helps strengthen cross-domain knowledge transfer and limits semantic drift or hallucinations in LLM outputs.


\subsubsection{Query Classification Component}

The hierarchical architecture balances computational efficiency with semantic precision through a two-stage process.

\textbf{Stage 1 (Lightweight Classification)}: We use the SetFit classifier~\cite{tunstall2022efficient}, fine-tuned on compact LLM models like Gemma 3B, which leverages few-shot techniques~\cite{Yehudai_2024}.
It optimizes embeddings using in-batch negatives and hard negative mining to achieve convergence with minimal labelled data.
We assess confidence using margin-based scoring and probability calibration.
High-confidence predictions proceed directly, while uncertain cases (e.g., near decision boundaries) escalate to Stage 2 to prevent error propagation.

For a query $q$, the component generates embeddings $e_q$ using sentence transformers, which yield logit scores and softmax-normalized probabilities.
Then, we calibrate confidence via the margin between top logits, and the following rule governs routing:
{\small
\begin{multline}
\text{Routing}_{\text{S1}} = \\
\begin{cases}
\text{Accept}                   & \text{confidence} \geq 0.85 \\
\text{Escalate to S2}           & 0.65 \leq \text{confidence} < 0.85 \\
\text{Request Clarification}    & \text{confidence} < 0.65
\end{cases}
\end{multline}
}
This strategy saves resources by skipping detailed investigation in high-confidence scenarios.

\noindent
\textbf{Stage 2 (LLM-Based Refinement)}: To dissect the decision process for escalated queries, we use chain-of-thought (CoT) reasoning~\cite{Wei_2022} to decompose the decision process.
It generates and applies task-specific prompts to highlight inference patterns and constraints, drawing upon information from the domain-aware component.
To achieve comprehensive coverage, we dynamically retrieve samples from unlabelled data using density-aware sampling to ensure broad coverage. 
On the other hand, alternative prompts elicit clarification inquiries in unclear instances (e.g., confidence 0.4 to 0.6).
The classification is output as the following structure:
\begin{multline}
    \text{Classification} = (\text{score} \in [0,1], \text{intent} \in \mathcal{I}, \text{context},\\
    \text{confidence} \in [0,1])
\end{multline}
where $\mathcal{I}$ denotes predefined intents, and context includes domain and temporal details. Algorithm~\ref{alg:query-classification} outlines the integrated process, enhancing accuracy for out-of-distribution queries while maintaining efficiency.

\begin{algorithm}[htb]
\caption{Two-Stage Query Classification}
\label{alg:query-classification}
\begin{algorithmic}[1]
\REQUIRE Query $q$, high threshold $\theta_{\text{high}}$, low threshold $\theta_{\text{low}}$, intent taxonomy $\mathcal{I}$, retrieval function $\text{Retrieve}(\cdot)$
\ENSURE Classification tuple, reflection feedback, escalation decision
\STATE $e_q \leftarrow \text{SentenceTransformer}(q)$
\STATE $\text{logits} \leftarrow \text{ClassifierHead}(e_q)$
\STATE $\text{probs} \leftarrow \text{softmax}(\text{logits})$
\STATE $\text{intent}^* \leftarrow \arg\max(\text{probs})$
\STATE $\text{score}^* \leftarrow \max(\text{probs})$
\STATE $\text{margin} \leftarrow \text{score}^* - \text{second\_max}(\text{probs})$
\STATE $\text{confidence}_{S1} \leftarrow \text{Calibrate}(\text{score}^*, \text{margin})$
\IF{$\text{confidence}_{S1} \geq \theta_{\text{high}}$}
    \STATE $\text{classification} \leftarrow (\text{score}^*, \text{intent}^*, \text{context}, \text{confidence}_{S1})$
    \STATE $\text{reflection} \leftarrow \text{GenerateReflection}(\text{classification}, C_{\text{classify}})$
    \STATE \textbf{return} $\text{classification}, \text{reflection}, \text{False}$
\ELSIF{$\text{confidence}_{S1} < \theta_{\text{low}}$}
    \STATE \textbf{return} $(\text{None}, \text{``Need more info — please clarify''}, \text{True})$
\ELSE
    \STATE $\mathcal{E} \leftarrow \text{Retrieve}(q, \text{diversity\_metric}, k=5)$
    \STATE $\text{Prompt} \leftarrow \text{ConstructCoTPrompt}(q, \mathcal{E}, \mathcal{I})$
    \STATE $\text{reasoning}, \text{intent}^{**}, \text{score}^{**} \leftarrow \text{LLM}(\text{Prompt})$
    \STATE $\text{confidence}_{S2} \leftarrow \text{ExtractConfidence}(\text{reasoning})$
    \STATE $\text{classification} \leftarrow (\text{score}^{**}, \text{intent}^{**}, \text{context}, \text{confidence}_{S2})$
    \STATE $\text{reflection} \leftarrow \text{GenerateReflection}(\text{classification}, C_{\text{classify}})$
    \STATE \textbf{return} $\text{classification}, \text{reflection}, \text{False}$
\ENDIF
\end{algorithmic}
\end{algorithm}

\subsubsection{Privacy Protection Component}

This component employs a context-aware anonymization mechanism to balance privacy with semantic preservation, which is important in root-cause analysis and troubleshooting. 
We apply anonymization at three criticality levels during self-reflection:
\begin{itemize}
    \item \textit{High-Criticality}: For entities such as network topologies, we employ structure-preserving methods to retain relational patterns for diagnosis.
    \item \textit{Medium-Criticality}: For entities like locations, we use $k$-anonymity-based generalizations~\cite{Sweeney_2002} to prevent unique identification.
    \item \textit{Low-Criticality}: We fully redact entities such as PII, as they offer no diagnostic value.
\end{itemize}
This tiered approach incorporates differential privacy via noise injection, minimizing information loss compared to uniform methods and defending against inference attacks.

Algorithm~\ref{alg:anonymization} formalizes this process~\cite{habernal2023privacy}, starting with entity recognition and criticality assessment through self-reflection.
We assign pseudonyms that preserve structure to high-criticality items, generalize medium-criticality items with Laplace noise for $\varepsilon$-differential privacy, and replace low-criticality items with placeholders.

\begin{algorithm}[htb]
\caption{Semantic-Preserving Anonymization}
\label{alg:anonymization}
\begin{algorithmic}[1]
\REQUIRE Query string $q$, PII entity types $\mathcal{E}$
\ENSURE Anonymized query $q'$, mapping $\mathcal{M}$
\STATE Initialize $\mathcal{M} \leftarrow \{\}$
\STATE Identify PII: $\{e_1, e_2, \ldots, e_n\} \leftarrow \text{NER}(q)$
\FOR{each entity $e_i$ with type $t_i$}
    \STATE $s_i \leftarrow \text{SemanticCriticality}(e_i, q, t_i)$
    \IF{$s_i = \text{HIGH}$}
        \STATE $e'_i \leftarrow \text{StructurePreserving}(e_i, t_i)$
        \STATE $\mathcal{M}[e'_i] \leftarrow e_i$
    \ELSIF{$s_i = \text{MEDIUM}$}
        \STATE $e'_i \leftarrow \text{CategoryReplacement}(e_i, t_i)$
    \ELSE
        \STATE $e'_i \leftarrow [\text{REDACTED}]$
    \ENDIF
    \STATE Replace $e_i$ with $e'_i$
\ENDFOR
\RETURN $q'$, $\mathcal{M}$
\end{algorithmic}
\end{algorithm}


\subsubsection{Query Translation Component}

Acting as a linguistic bridge, this component converts anonymized user queries into precise technical formats for specialized LLMs.
Algorithm~\ref{alg:query-translation} provides a detailed implementation demonstration.
We use semantic decomposition to identify diagnostic intents \cite{zhou2025llm}, extract constraints (e.g., network segments), and map terminology to domain ontologies.
In addition, context enrichment from the domain-aware component allows us to infer relevant factors, such as latency needs in smart grids.
Furthermore, the component refines the query using reasoning and self-reflection techniques to make it align with the requirements for technical and constraint validity.
This method generates consistent queries with the specialist LLM, decreasing ambiguity and increasing diagnostic accuracy.

\begin{algorithm}[htb]
\caption{Query Translation}
\label{alg:query-translation}
\begin{algorithmic}[1]
\REQUIRE Anonymized query $q_{\text{anon}}$, domain context $\text{ctx}_{\text{domain}}$, ontology $\mathcal{O}_{\text{telecom}}$
\ENSURE Technical query $q_{\text{tech}}$, constraints $\mathcal{C}_{\text{tech}}$, confidence $\text{conf}_{\text{trans}}$

\STATE $\text{intent} \leftarrow \text{IntentClassifier}(q_{\text{anon}})$
\STATE $\mathcal{C}_{\text{explicit}} \leftarrow \text{ConstraintExtractor}(q_{\text{anon}})$

\STATE $\mathcal{C}_{\text{implicit}} \leftarrow \text{InferConstraints}(\text{ctx}_{\text{domain}}, \mathcal{O}_{\text{telecom}})$
\STATE $\mathcal{C}_{\text{all}} \leftarrow \mathcal{C}_{\text{explicit}} \cup \mathcal{C}_{\text{implicit}}$

\FOR{each user term $t_{\text{user}}$ in $q_{\text{anon}}$}
    \STATE $t_{\text{tech}} \leftarrow \text{OntologyLookup}(t_{\text{user}}, \mathcal{O}_{\text{telecom}})$
\ENDFOR

\STATE $q_{\text{intermediate}} \leftarrow \text{FormulateQuery}(q_{\text{anon}}, \mathcal{C}_{\text{all}}, \mathcal{O}_{\text{telecom}})$
\STATE $\text{reflection} \leftarrow \text{Reflect}(q_{\text{intermediate}}, \mathcal{C}_{\text{checklist}})$

\IF{$\text{reflection.issues} > 0$}
    \STATE $q_{\text{intermediate}} \leftarrow \text{RefineQuery}(q_{\text{intermediate}}, \text{reflection})$
\ENDIF

\STATE $\text{conf}_{\text{trans}} \leftarrow \text{ValidateTranslation}(q_{\text{intermediate}}, q_{\text{anon}})$

\STATE $q_{\text{tech}} \leftarrow q_{\text{intermediate}}$
\STATE $\mathcal{C}_{\text{tech}} \leftarrow \text{FormalizeConstraints}(\mathcal{C}_{\text{all}})$
\RETURN $q_{\text{tech}}, \mathcal{C}_{\text{tech}}, \text{conf}_{\text{trans}}$
\end{algorithmic}
\end{algorithm}


\subsubsection{Response Simplification Component}

This component converts technical replies to plain language while maintaining diagnostic consistency.
It employs multi-pass processing, recognizes jargon, substitutes simplifications, and uses self-reflection to evaluate.
We evaluate readability using Flesch Reading Ease (FRE) metrics ~\cite{marulli2024understanding} to measure the readability.
Algorithm~\ref{alg:simplification} details the pipeline of this component.
First, we extract concepts and actions, map them via a lexicon refined through self-reflection, and verify the output against thresholds to avoid over-simplification.
We keep quantitative elements unchanged to preserve technical accuracy.
Finally, we use an on-premise LLM-generated mapping (see Table~\ref{tab:term_mappings}) to convert technical terms into user-friendly expressions.

\begin{algorithm}[htb]
\caption{Technical Response Simplification}
\label{alg:simplification}
\begin{algorithmic}[1]
\REQUIRE Technical response $r_{\text{tech}}$, user query $p$
\ENSURE Simplified response $r_{\text{user}}$
\STATE Extract concepts: $\mathcal{C} \leftarrow \text{ConceptExtractor}(r_{\text{tech}})$
\STATE Build mappings: $\mathcal{T}$
\STATE Identify actions: $\mathcal{A} \leftarrow \text{ActionExtractor}(r_{\text{tech}})$
\FOR{each action $a \in \mathcal{A}$}
    \STATE $a' \leftarrow \text{SimplifyAction}(a, p, \mathcal{T})$
    \STATE Append $a'$ to $r_{\text{user}}$
\ENDFOR
\STATE Generate verification prompt
\STATE check $\leftarrow \text{LLM}(v \oplus r_{\text{user}})$
\IF{check.confidence $< \theta_{\text{verify}}$}
    \STATE Regenerate with adjusted complexity
\ENDIF
\RETURN $r_{\text{user}}$
\end{algorithmic}
\end{algorithm}

\begin{table}[htb]
    \centering
    \begin{tabular}{|l|l|}
    \hline
    \textbf{Technical Term} & \textbf{User-Friendly Version} \\
    \hline
    RSRP (Reference Signal Received Power) & Signal Strength \\
    SINR (Signal-to-Interference-plus-Noise Ratio) & Signal Quality \\
    RLC Retransmission & Data Re-sending \\
    PRB Utilization & Network Congestion \\
    Cell Reselection & Tower Switching \\
    Backhaul Congestion & Overloaded Links \\
    DU Latency Spike & Processing Delay \\
    \hline
    \end{tabular}
    \caption{LLM-Generated Mappings from Telecom Terms to Plain English Examples}
    \label{tab:term_mappings}
\end{table}

\subsubsection{Reasoning and Self-Reflection Component}

Integrated with the reasoning model, this component uses self-reflection to check the output quality at each pipeline stage. Based on the Reflexion and Self-Refine frameworks~\cite{Shinn_2023, Madaan_2023}, it generates written feedback by comparing the output against checklists and user queries.
We then use prompts to get feedback on semantic correctness, precision, and alignment, and we save these results for future improvement.

Building on the ReAct paradigm~\cite{Yao_2023}, multi-agent coordination incorporates reflection at decision points:
\begin{multline}
\text{Agent Output}_{t+1} = \\
f(\text{Action}_{t}, \text{Observation}_{t}, \text{Reflection}_{t})
\end{multline}

Reflections are produced as:
\begin{equation}
\text{Reflection}_{t} = \text{LLM}(\text{"Evaluate output"} \oplus \text{CheckList}_{agent})
\end{equation}

\begin{table}[htb]
\centering
\caption{Component-Specific Self-Reflection Checklists}
\label{tab:reflection-checklists}
\renewcommand{\arraystretch}{1.3}
    \rowcolors{2}{white}{gray!5}
    \begin{tabular}{p{1.3cm} p{7cm}}
    \toprule
    \textbf{Component} & \textbf{Self-Reflection Checklist Items} \\
    \midrule
    \textbf{Domain-Aware} &
    \begin{itemize}[leftmargin=*, label=$\bullet$, itemsep=0.8pt, parsep=0pt, topsep=2pt]
        \item Probability distribution validity and entropy calibration
        \item Confidence score alignment with domain assignment likelihood
        \item Coverage of all relevant industrial verticals
        \item Feature embedding coherence with domain ontologies
        \item Absence of competing domain interpretations
        \item Prompt template relevance to the identified domain
        \item Hallucination risk assessment
    \end{itemize} \\
    \midrule
    \textbf{Query Classification} &
    \begin{itemize}[leftmargin=*, label=$\bullet$, itemsep=0.8pt, parsep=0pt, topsep=2pt]
        \item Margin-based confidence score calibration
        \item Decision boundary proximity assessment
        \item Alternative classification interpretations considered
        \item Escalation threshold appropriateness
        \item Feature space representation completeness
        \item Exemplar diversity and semantic coverage (Stage 2)
        \item Chain-of-thought reasoning transparency
    \end{itemize} \\
    \midrule
    \textbf{Privacy Protection} &
    \begin{itemize}[leftmargin=*, label=$\bullet$, itemsep=0.8pt, parsep=0pt, topsep=2pt]
        \item Comprehensive PII entity detection and classification
        \item Criticality level assignment accuracy
        \item Residual re-identification vulnerability assessment
        \item k-anonymity group size validation
        \item Structure preservation fidelity for network topology
        \item Noise injection parameter appropriateness
        \item Differential privacy guarantee verification
    \end{itemize} \\
    \midrule
    \textbf{Query Translation} &
    \begin{itemize}[leftmargin=*, label=$\bullet$, itemsep=0.8pt, parsep=0pt, topsep=2pt]
        \item Diagnostic intent extraction completeness
        \item Temporal and spatial constraint identification
        \item User terminology mapping to technical nomenclature
        \item Domain context integration accuracy
        \item Implicit parameter inference validation
        \item Technical specification precision and unambiguity
        \item Translation confidence metadata calibration
    \end{itemize} \\
    \midrule
    \textbf{Response Simplification} &
    \begin{itemize}[leftmargin=*, label=$\bullet$, itemsep=0.8pt, parsep=0pt, topsep=2pt]
        \item Technical accuracy preservation versus original
        \item Quantitative finding retention without modification
        \item Lexicon mapping appropriateness and consistency
        \item Readability metric compliance (Flesch-Kincaid 7--9)
        \item Semantic ambiguity absence
        \item Domain expert information integrity retention
        \item Cognitive accessibility for non-specialists
    \end{itemize} \\
    \bottomrule
    \end{tabular}
\end{table}

We use the component-specific checklists in Table~\ref{tab:reflection-checklists} to ensure targeted assessments, which foster consistent quality across the system.
Please keep in mind that, in order to improve the accuracy and eliminate bias in self-reflection, the critic model that provides feedback is another LLM agent.


\section{Evaluation Proposal Framework} \label{sec:evaluation}


\subsection{Validation Dataset}

We construct the validation dataset through a systematic approach that combines existing telecommunications question-answer pairs with domain-specific augmentation.
The dataset generation follows three stages: base data selection, vertical domain sampling, and intent diversification.
In detail, TeleQnA~\cite{Maatouk_2025} is used as the foundation for base question and expected answer pairs.
This baseline ensures that the generated data is consistent with prior telecommunications datasets and reduces bias in problem formulation.


\begin{figure*}[htbp]
    \centering
    \begin{minipage}[c]{0.58\linewidth}
        \centering
        \begin{subfigure}[b]{\linewidth}
            \centering
            \includegraphics[width=\linewidth]{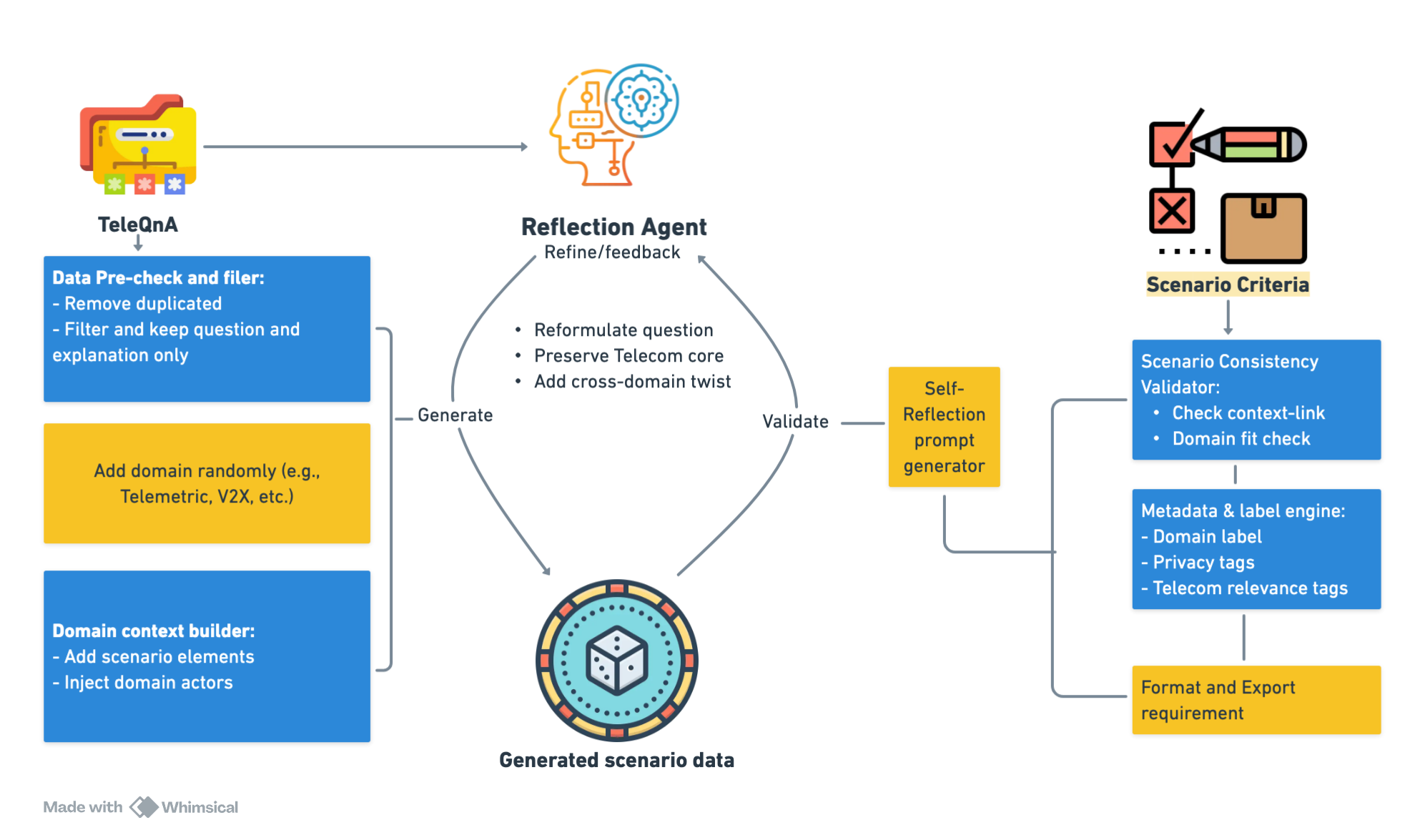}
            \caption{Self-Reflection data generator.}
            \label{fig:data-generator}
        \end{subfigure}
    \end{minipage}
    \hfill 
    \begin{minipage}[c]{0.33\linewidth}
        \centering
        \begin{subfigure}[b]{\linewidth}
            \centering
            \includegraphics[width=\linewidth]{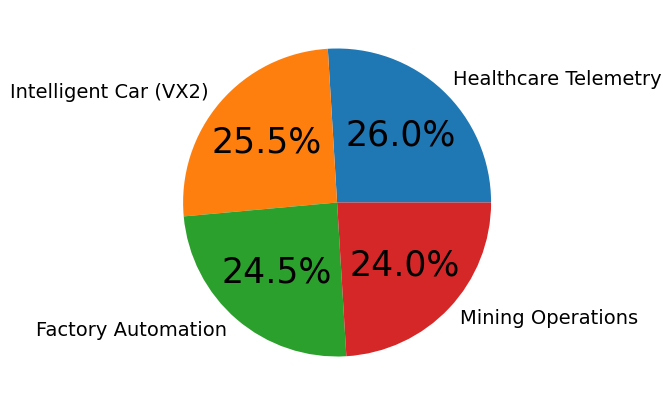}
            \caption{Vertical industrial domains.}
            \label{fig:vertical-domain-distribution}
        \end{subfigure}
        
        \vspace{0em} 
        
        \begin{subfigure}[b]{\linewidth}
            \centering
            \includegraphics[width=\linewidth]{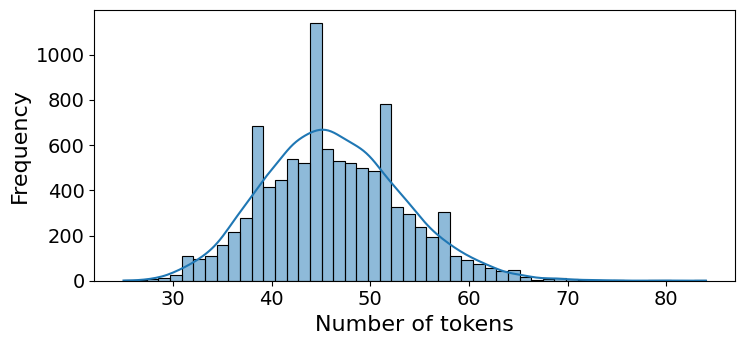}
            \caption{Token length.}
            \label{fig:distribution-token-length}
        \end{subfigure}
    \end{minipage}

    \caption{Dataset overview: (a) The data generation pipeline, (b) Distribution of vertical industrial domains, and (c) Distribution of token length.}
    \label{fig:dataset_combined_view}
\end{figure*}

\begin{figure}[htb]
    \centering
    \includegraphics[width=1\linewidth]{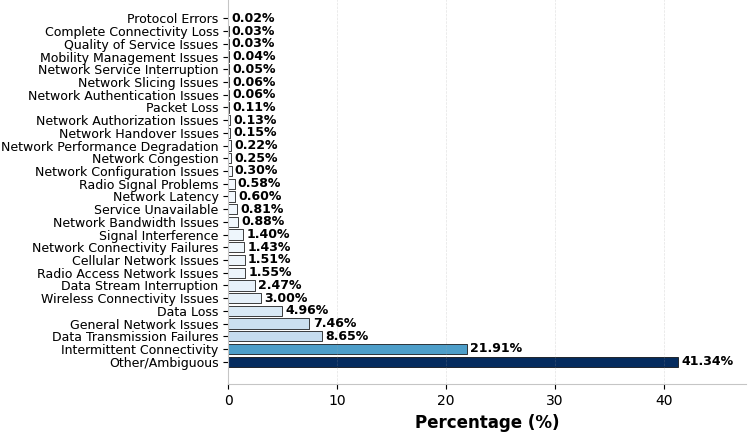}
    \caption{Intent categories.}
    \label{fig:dataset_category_distribution}
\end{figure}

To generate samples specific to vertical domains, we apply a self-reflection method to randomly select from a predefined set of industrial applications, as demonstrated in the Fig.~\ref{fig:data-generator}.
The fields of study chosen include healthcare telemetry, vehicular communication (V2X), mining operations, and factory automation, which may be altered depending on the validation situation and distribution.
Fig. \ref{fig:vertical-domain-distribution} illustrates the vertical domain distribution in our dataset.
For each base question from TeleQnA, we contextualize it within the selected vertical by introducing domain-specific constraints and requirements through self-reflection prompt generation.
This approach maintains the core problem structure while incorporating domain-relevant context.



As a consequence, the generated dataset consists of 10,000 validation scenarios.
The dataset and prompts used in this paper are available in the accompanying project repository~\footnote{Dataset and prompts URL: \url{https://ericsson-mtl.gitlab.io/cross-domain-query-translation-for-llm-based-telecom}}.
Furthermore, the intent distribution shows the natural frequency of issues encountered in telecommunications systems, with intermittent connectivity and data transmission failures accounting for the majority of the dataset, while rare scenarios such as packet loss and network authentication issues are also included.
This distribution provides that the model will come across both common failure modes and edge cases during validation.
Fig.~\ref{fig:dataset_category_distribution} provides details on intent categories and their respective frequencies within our dataset.
In addition, Fig.~\ref{fig:distribution-token-length} illustrates the distribution of query token length.
The dataset ranges from simple, short queries that describe basic connection concerns to complicated, lengthy questions that include many network levels, various failure symptoms, and domain-specific terms.
This query variation evaluates if the model performs consistently for both clear, well-articulated issue statements and verbose or ambiguous real-world scenarios.

\subsection{Evaluation method}

We adopt a hybrid evaluation strategy that combines deterministic statistical metrics with an LLM-as-a-judge framework.
For the domain-aware classification and privacy-protection modules, we report accuracy, precision, recall, and F1 score. 
Privacy evaluation additionally balances anonymization strength with diagnostic utility through three metrics: PII Recall (correctly masked sensitive entities), Token Retention Rate (non-sensitive content preserved), and the Preservation Score (semantic similarity between original and anonymized queries).

Since cross-domain translation lacks explicit ground truth, an LLM-as-a-judge is used to provide qualitative assessment.
In particular, we employed the TSLAM model\footnote{\url{https://huggingface.co/NetoAISolutions/TSLAM}}, which is reported in the GSMA Open-Telco LLM Benchmarks\footnote{\url{https://huggingface.co/spaces/otellm/leaderboard}}.
We selected TSLAM due to its design and fine-tuned it for telecommunications and its ability to be deployed locally within our internal system.
Following the Critic Model architecture~\cite{Wei_2025}, a telecom-specialized LLM evaluates outputs using the component-specific checklists in Table~\ref{tab:reflection-checklists}.
The judge measures hallucination rate, constraints introduced without user evidence, and diagnostic consistency, verifying that the translated technical query maintains the reasoning implied by the original user statement.
Table~\ref{tab:e2e-example} presents a simplified end-to-end example using our framework.

\begin{table*}[ht]
    \centering
    \small 
    \caption{An Example of End-to-End Cross-Domain Processing Pipeline}
    \label{tab:pipeline_example}
    \renewcommand{\arraystretch}{1.2}
    \begin{tabular}{p{0.22\linewidth} p{0.73\linewidth}}
    \toprule
    \textbf{Pipeline Stage} & \textbf{Agent Output / Action} \\
    \midrule

    \textbf{User Input (or Domain X LLM)} & 
    \textit{"Dr. Ramirez reports real-time heart rate data from patient John Smith's wearable (ID: WM-47B-22, IP: 10.24.1.15) stopped updating in the ICU."} \\
    \midrule

    \textbf{1. Domain-Aware} & 
    \textbf{Vertical:} Healthcare Telemetry (Confidence: 0.96) \newline
    \textbf{Context:} Remote Patient Monitoring / Latency-Critical. \\
    \midrule

    \textbf{2. Query Classification} \newline
    (Algorithm 1) & 
    \textbf{Decision:} \textcolor{blue}{Stage 1 Fast-Path} (SetFit Score: 0.92 $> \theta_{high}$). \newline
    \textbf{Intent:} \texttt{Network\_Connectivity\_Loss}. \newline
    \textbf{Routing:} Valid Telecom Fault $\rightarrow$ Proceed to Privacy Filter. \\
    \midrule

    \textbf{3. Privacy Agent} \newline
    (Algorithm 2) & 
    \textbf{Action:} High-Criticality Preservation (Topology); Low-Criticality Redaction. \newline
    \textbf{Output:} "Physician reports telemetry stopped from [DEVICE\_WEARABLE] (IP: [IP\_PATIENT]) at [REDACTED\_LOC]. Patient is post-op." \\
    \midrule

    \textbf{4. Translation Agent} \newline
    (Algorithm 3) & 
    \textbf{Tech Intent:} \texttt{telemetry\_interruption}, \texttt{MFAF\_data\_mapping}, \texttt{iot\_connectivity}. \newline
    \textbf{Translated Query:} "Diagnose cause of data stream termination for IoT device [IP\_PATIENT] in Core Network. Check MFAF configuration for outbound notification mapping failures." \\
    \midrule
    
    \textbf{5. Telecom Expert} \newline
    (Specialized LLM) & 
    \textbf{Technical Diagnosis:} The issue matches the \texttt{Nmfaf\_3daDataManagement\_Deconfigure} operation (3GPP Rel-18). The MFAF has terminated the mapping between the analytics stream and the notification endpoint, halting outbound data flow. \\
    \midrule

    \textbf{6. Simplification} \newline
    (Algorithm 4)  & 
    \textbf{User-Friendly Response:} "The patient monitor stopped updating because a 'Deconfigure' command disconnected the data stream from the central dashboard. This is a configuration setting in the network (MFAF), not a failure of the device itself." \newline
    \textit{Readability Score (FRE): 65.2 (Standard)} \\
    \bottomrule
    \end{tabular}
    \label{tab:e2e-example}
\end{table*}
\subsection{Evaluation}

\begin{table}[htb]
\caption{Domain-aware performance}
\centering
\begin{tabular}{|l|c|c|}
\hline
\textbf{Industry} & \textbf{F1} & \textbf{Accuracy} \\
\hline
Healthcare & 0.93 & 0.91 \\
Factory & 0.96 & 0.94 \\
Smart Cars & 0.91 & 0.89 \\
Mining Operations & 0.93 & 0.92 \\
\hline
\textbf{Overall} & \textbf{0.95} & \textbf{0.92} \\
\hline
\end{tabular}
\label{tab:domain_aware_performance}
\end{table}

Reported in Table~\ref{tab:domain_aware_performance}, the Domain-Aware component demonstrated strong robustness across all industrial verticals, achieving an overall F1 score of 0.95 and an accuracy of 0.927, confirming the effectiveness of the soft classification strategy.
Factory Automation attained the highest performance (F1 = 0.96), attributed to the sector’s well-defined communication terminology, while Smart Cars showed slightly lower performance (F1 = 0.91), likely due to overlapping vocabulary with general enterprise Wi-Fi scenarios, which contributed to minor classification ambiguity.

\begin{table}[htbp]
\centering
    \begin{tabular}{lcc}
    \hline
    \textbf{Metric} & \textbf{With Self-Reflection} & \textbf{Without Self-Reflection} \\
    \hline
    PII Recall & 93.25\% & 93.25\% \\
    Precision & 1 & 1 \\
    F1 Score & 0.96 & 0.96 \\
    Preservation Score & 80.16 / 100 & 84.65 / 100 \\
    Token Retention & 89.86\% & 90.00\% \\
    \hline
    \end{tabular}
    \caption{Privacy Protection Validation Results}
    \label{tab:privacy_protection_res}
\end{table}

We conducted an ablation study to assess the effect of the self-reflection (LLM-as-a-judge) mechanism on privacy enforcement, with results summarized in Table~\ref{tab:privacy_protection_res}. Both the baseline and reflection-augmented models achieved identical PII recall (93.25\%) and perfect precision (1.0), indicating consistent detection and masking of sensitive entities.
However, the reflection-enabled variant produced a lower preservation score (80.16 vs. 84.65), reflecting stricter anonymization behavior.
This reduction is expected, as the self-reflection loop generalizes medium-criticality entities through $k$-anonymity rather than retaining their raw form, thereby enforcing a more conservative privacy posture aligned with high-criticality handling requirements.
At the same time, the higher token-retention score suggests that the reflection-enabled model preserves more non-sensitive contextual tokens, which helps maintain the structural coherence of the anonymized query despite stricter semantic generalization.

\begin{table}[htbp]
    \centering
    \begin{tabular}{lc}
        \toprule
        \textbf{Metric} & \textbf{Score (/100)} \\
        \midrule
        Average Semantic Overlap & 79.24 \\
        Average Coverage & 73.10 \\
        Average Similarity & 79.37 \\
        Average Hallucination Rate & 13.61 \\
        \bottomrule
    \end{tabular}
    \caption{Query-to-technical translation performance.}
    \label{tab:query_translation_performance_eval}
\end{table}

The results in Table~\ref{tab:query_translation_performance_eval} indicate that the query-to-technical translation component achieves strong semantic alignment, with an average semantic overlap of 79.24 and similarity of 79.37, suggesting that the technical reformulation preserves most of the user's intent and contextual meaning.
The coverage score of 73.10 shows that the model captures most required technical attributes, although some implicit parameters remain omitted in complex diagnostic scenarios.
Specifically, the hallucination rate remains low at 13.61, indicating that the model rarely introduces technical constraints not supported by the original query within the target vertical domain.

\begin{table}[htbp]
    \centering
    \begin{tabular}{lc}
        \toprule
        \textbf{Metric} & \textbf{Score (/100)} \\
        \midrule
        Average Flesch Reading Ease & 55.31 \\
        Average Domain Appropriateness & 78.59 \\
        Average Clarity \& Simplicity & 74.13 \\
        Average Completeness & 75.43 \\
        Average Actionability & 73.89 \\
        \midrule
        \textbf{Average Overall Readability} & \textbf{74.67} \\
        \bottomrule
    \end{tabular}
    \caption{Readability Evaluation of Generated Responses}
    \label{tab:readability_eval}
\end{table}

The readability results in Table~\ref{tab:readability_eval} show that the generated responses achieve an overall readability score of 74.67, indicating that the system consistently produces explanations suitable for non-technical users.
The domain-appropriateness score of 78.59 confirms that responses remain aligned with telecommunications concepts while avoiding unnecessary jargon.
Clarity and Simplicity (74.13) and Completeness (75.43) further demonstrate that the reformulated answers preserve essential diagnostic information without overwhelming users. 
The actionability score of 73.89 indicates that the model provides guidance that is sufficiently concrete for users to follow.
Although the FRE score (55.31) reflects moderate linguistic complexity, the combined metrics show that the system effectively balances technical accuracy with user-friendly communication.
This score is expected, as the system must retain essential terminology from both telecommunications and the target vertical domain (e.g., healthcare), making full simplification infeasible without losing diagnostic precision.


\section{Conclusion}
\label{sec:discussion}

This paper introduced a cross-domain query translation framework that integrates multi-agent LLM coordination, hierarchical classification, semantic-preserving anonymization, and reflection-driven reasoning to bridge the gap between non-technical user inputs and telecom diagnostics.
Our evaluation across 10,000 synthetic validation scenarios demonstrates strong domain classification accuracy, faithful technical translation with low hallucination rates, and readable user-facing responses while maintaining privacy constraints.

Future work will focus on expanding empirical validation with real-world telecom data, incorporating human expert assessments, and calibrating the LLM-as-a-judge with multi-critic or ensemble verification.
We also plan to benchmark against RAG-based and domain-adapted telecom LLM systems, and to strengthen the privacy mechanism with formal differential privacy guarantees.

\section*{Acknowledgment}

The first author was supported by an internship at Ericsson in Montréal, Québec, Canada.
We would like to thank Yiwei Wen and Bo Yang for their valuable support throughout this project.

\bibliographystyle{IEEEtran}
\bibliography{references}



\end{document}